# A Stability Analysis for the Reaction Torque Observer-based Sensorless Force Control Systems

Emre Sariyildiz
School of Mechanical, Materials, Mechatronic, and Biomedical Engineering,
Faculty of Engineering and Information Sciences
University of Wollongong
Wollongong, NSW, 2522, Australia
emre@uow.edu.au

*Abstract*— **This paper proposes a new stability analysis for the Reaction Torque Observer (RTOb) based robust force control systems in the discrete-time domain. The robust force controller is implemented by employing a Disturbance Observer (DOb) to suppress disturbances, such as friction and hysteresis, in an inner-loop and another disturbance observer, viz RTOb, to estimate contact forces without using a force sensor. Since the RTOb-based robust force controllers are always implemented using computers and/or microcontrollers, this paper proposes a stability analysis in the discrete-time domain. It is shown that the bandwidth of the DOb is limited not only by the noise of velocity measurement but also by the waterbed effect. It is also shown that the stability of the robust force controller may significantly deteriorate when the design parameters of the RTOb are not properly tuned. For example, the robust force controller may have a non-minimum phase zero(s) as the design parameter of the identified inertia (torque coefficient) of the RTOb is increased (decreased). This may lead to poor stability and performance in force control applications. The proposed stability analysis conducted in the discrete-time domain is verified by simulations and experiments.**

## I. INTRODUCTION

Compared to traditional industrial robotic manipulators, next-generation robotic systems (e.g., exoskeletons, prostheses, humanoids, and collaborative and medical robots) are expected to physically interact with unknown, dynamic and unstructured environments such as human beings [1 – 6]. When it comes to physical robot-environment interaction, it is a well-known fact that precise positioning is not sufficient [1, 7]. To safely perform physical interaction tasks, it is essential to precisely control the contact force between robot and environment [6 – 8]. Although various direct and indirect force controllers have been proposed in the last decades, physical robot environment interaction tasks still suffer from the stability and low-performance problems of force control systems, particularly when the robots perform tasks in unstructured environments [6 – 15].

To improve the stability and performance of physical robot-environment interaction tasks, the RTOb-based robust force controller was proposed by Murakami and Ohnishi in 1993 [16]. Since then it has been applied to various robotic applications spaning from industrial and rehabilitation robotics to automotive and medical robotics [2 – 4, 17 – 19]. An RTOb-based robust force controller is synthesised by employing two DObs in an inner- and an outer- loop [16]. While the robustness of the force controller is improved by suppressing disturbances via the DOb in the inner-loop, the contact force between robot and environment is estimated using the RTOb in the outer-loop. This force-sensorless force control technique provides several benefits in practice. For example, i) contact forces can be directly estimated without changing the compliance of the force control system [7], ii) compared to a force sensor, a higher bandwidth of contact force estimation can be achieved using the RTOb [20], and iii) the force-sensorless force control technique not only reduces the size and mechanical complexity of force control systems but also enables low-cost physical robot-environment interaction applications [19]. Nevertheless, the RTOb is a model-based force controller so the dynamic model of the motion control systems should be identified to achieve good stability and performance in practice [16].

In the literature, it is generally assumed that the RTOb-based robust force controller is designed by using the actual inertia and torque coefficient values of servo systems [12, 21]. However, this conventional design approach could be impractical in many applications because we may not obtain an accurate dynamic model for some robotic systems such as wearable devices [12]. It is therefore essential to understand how the stability and performance of physical interaction tasks change by the design parameters of the robust force controller. Moreover, the conventional design approach does not allow us to adjust the stability and performance of the robust force controller by tuning the design parameters of the DOb and RTOb.

Several studies have been conducted to improve the stability and performance of the RTOb-based robust force controller in the last two decades. For example, an accelerometer was integrated to the design of the RTOb so that the bandwidth of force estimation was increased in PAIDO [22]. It is shown in [22] that increasing the bandwidth of the RTOb improves not only the accuracy of force estimation but also the stability of contact motion. To further increase the bandwidth of force estimation, Kalman filter was integrated to the RTOb in [23, 24]. However, the robust stability and performance of the force controller are not considered in these studies. To understand how the design parameters of the DOb and RTOb affect the robust force controller, different stability and performance analyses have been proposed. For example, [25] and [26] show that the stability of the robust motion controller can be adjusted by tuning the design parameters of the DOb in the inner-loop. [7] shows that not only the DOb but also the RTOb can be used to tune the stability and performance of the robust force controller. Although these analyses provide good insight into the stability and performance of the robust force controller, they fall-short in explaining some dynamic responses in practice because they are conducted in the continuous-time domain. For example, continuous-time analyses cannot explain why the robust force controller becomes unstable as the bandwidth of the DOb is increased in the inner-loop [27, 28]. Since the DOb and RTOb are always implemented using computers and microcontrollers, it is essential to understand the stability and performance of the RTOb-based digital robust force controllers by conducting an analysis in the discrete-time domain [29 – 32].

To this end, a new stability analysis is proposed for the RTOb-based digital robust force controllers in this paper. To derive the design constraints of the DOb in the inner-loop, the discrete Bode-Integral Theorem is employed. This theorem shows that the robust stability and performance of the inner-loop controller deteriorate due to the waterbed effect as the bandwidth of the digital DOb increases. To tackle this problem, the sampling-time of the robust force controller should be decreased. This, however, generally increases cost in practical engineering applications. The proposed stability analysis also shows that the design parameters of the RTOb can notably change the stability and performance of the robust force controller. As the identified inertia (torque coefficient) is increased (decreased) in the design of the RTOb, an open-loop zero moves towards the out of the unit circle. In other words, the robust force controller may have a non-minimum phase zero. This leads to a strict design constraint on the force control gain and may notably deteriorate the stability of the robust force controller. The stability and performance of the digital robust force controller can also be adjusted by tuning the bandwidths of the DOb and RTOb. While the robust force controller has a phase-lead controller when the bandwidth of the RTOb is larger than that of DOb, increasing the bandwidth of the DOb may result in a phase-lag compensator. The dynamic response of the digital robust force controller can be adjusted by tuning the phase- lead/lag compensator. The proposed stability analysis is verified by presenting simulation and experimental results.

The rest of the paper is organised as follows. Section II briefly introduces the DOb and RTOb in the discrete-time domain. Section III introduces the RTOb-based robust force controller and proposes a new stability analysis. Section IV verifies the proposed stability analysis by simulations and experiments. The paper ends with conclusion in Section V.

## II. DESIGN AND IMPLEMENTATION OF THE DOB AND RTOB IN THE DISCRETE-TIME DOMAIN

In this section, the design and implementation of the digital DOb and RTOb are briefly introduced.

### A. Disturbance Observer

The block diagram of the digital DOb implemented by the Backward-Euler integration method is illustrated in Fig. 1. The internal and external disturbances (e.g., friction, hysteresis and plant uncertainties) are estimated using the nominal plant model, control signal and velocity measurement. The robustness of the servo system is simply attained by feedbacking the estimated disturbances as shown in Fig. 1. In this figure, $J_m$ and $J_{m_n}$ are the actual and nominal inertiae, respectively; $K_\tau$ and $K_{\tau_n}$ are the actual and nominal thrust coefficients, respectively; $\tau_c$ and $\eta$ are the external load and noise, respectively; $q$, $\dot{q}$ and $\ddot{q}$ are the angle, velocity and acceleration, respectively; $g_{DOb}$ is the bandwidth of the DOb; $I$ is the current of the DC motor; $\tau_{dis}$ and $I_{dis}$ are the fictitious disturbance torque and current variables, respectively; $ZoH$ is the Zero order Hold; $T_s$ is the sampling-time; $t$ and $t_k = kT_s$ represent time in the continuous and discrete domains, respectively; $s$ and $z = e^{sT_s}$ are complex variables; $\hat{\bullet}$ is the estimation of $\bullet$; and $\bullet_{des}$ is the desired $\bullet$.

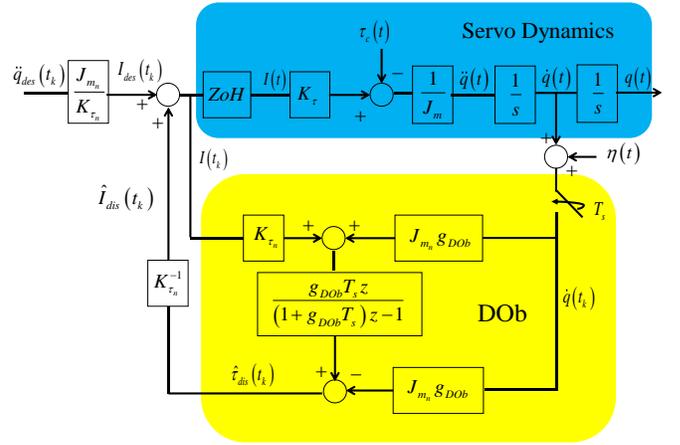

Fig. 1: Block diagram of the digital DOb implemented by the Backward-Euler Integration method.

The transfer function between the exogenous inputs $\ddot{q}_{des}(z)$, $\tau_c(z)$ and $\eta(z)$ to the output $\ddot{q}(z)$ can be directly derived from Fig. 1 as follows:

$$\ddot{q}(z) = \alpha \frac{(1+g_{DOb}T_s)z-1}{z-(1-\alpha g_{DOb}T_s)}\ddot{q}_{des}(z) - \frac{1}{J_m}S(z)\tau_c(z) - \frac{(z-1)}{T_s}T(z)\eta(z) \quad (1)$$

where $S(z) = \dfrac{z-1}{z-(1-\alpha g_{DOb}T_s)}$ and $T(z) = \dfrac{\alpha g_{DOb}T_s}{z-(1-\alpha g_{DOb}T_s)}$ are the discrete sensitivity and complementary sensitivity transfer functions in which $\alpha = \left(J_{m_n}K_\tau\right)/\left(J_m K_{\tau_n}\right)$.

Equation (1) shows that a phase-lead/lag compensator is synthesised in the inner-loop of the DOb-based robust motion control system. The phase margin of the digital controller improves as $\alpha$ is increased (i.e., the nominal inertia/thrust coefficient is increased/decreased). However, the phase margin of the inner-loop is limited by the stability constraint of the digital robust motion controller. As shown in Eq. (1), the inner-loop controller exhibits oscillatory response when $\alpha g_{DOb}T_s > 1$ and becomes unstable when $\alpha g_{DOb}T_s > 2$. To improve the bandwidth of disturbance estimation and phase margin in the inner-loop, the sampling-time of the digital robust motion controller should be decreased.

Equation 1 also shows that as $\alpha$ and $g_{DOb}$ are increased, while the sensitivity function becomes smaller at low frequencies the complementary sensitivity function gets larger values at higher frequencies. In other words, the robustness against disturbances is improved at low-frequencies but the digital motion controller becomes more sensitive to noise. Therefore, the design parameters of the DOb are limited by the noise and stability constraints of the digital motion controller.

Another constraint on the design parameters of the DOb can be obtained by applying the discrete Bode Integral Theorem to the inner-loop as follows:

$$\int_{-\pi}^{\pi} \ln \left| S\left(e^{j\omega T_s}\right) \right| d\omega T_s = -2\pi \ln \left|1 + \lim_{z \to \infty} L(z)\right| = 0 \quad (2)$$

where $L(z) = \dfrac{\alpha g_{DOb} T_s}{z-1}$ is the open-loop transfer function of the digital robust motion controller illustrated in Fig. 1 [33 – 36].

Equation (2) shows that the DOb-based digital robust motion controller suffers from the waterbed effect. To satisfy Eq. (2), the peak of the sensitivity function increases at high frequencies as it is decreased by using the higher values of $\alpha$ and/or $g_{DOb}$ to suppress disturbances at low frequencies. In other words, the design parameters of the DOb are constrained by the waterbed effect. Violating this design constraint may lead to severe stability and performance issues in practice. To tune the digital DOb-based robust motion controller, a design tool can be developed using Eq. (2), e.g., [29, 30].

*B. Reaction Torque Observer*

The block diagram of the digital RTOb implemented by the Backward-Euler integration method is illustrated in Fig. 2. In this figure, $J_{mi}$ and $K_{\tau i}$ are the identified inertia and torque coefficient, respectively; and $\tau_{di}$ is the identified disturbances due to the internal dynamics of the servo system such as gravity and friction. The other parameters are same as defined earlier.

As shown in Fig. 2, the contact force $\tau_c$ can be similarly estimated by using the identified inertia, torque coefficient and internal dynamics in the design of the observer. Since the RTOb is a model-based controller, the mismatch between the actual and identified plant models directly contributes to the error of contact force estimation as shown in Eqs. (3) and (4).

$$\tau_c(t_k) = K_i I(t_k) - J_i \ddot{q}(t_k) - \tau_d(t_k) \quad (3)$$

$$\begin{aligned}\hat{\tau}_c(z) &= \left(K_{\tau i} I(z) + J_i g_{RTOb} \dot{q}(z) - \tau_{di}(z)\right) \dfrac{g_{RTOb} T_s z}{(1+g_{RTOb} T_s)z - 1} - J_i g_{RTOb} \dot{q}(z) \\ &= \tau_c(z) \dfrac{g_{RTOb} T_s z}{(1+g_{RTOb} T_s)z - 1} - \left(\tau_{di}(z) - \tau_d(z)\right) \dfrac{g_{RTOb} T_s z}{(1+g_{RTOb} T_s)z - 1}\end{aligned} \quad (4)$$

where $\tau_d$ represents a fictitious disturbance variable that includes unknown and/or unmodelled disturbances of the servo system, and $\tau_{d_i}$ represents the identified $\tau_d$ that is used in the design of the RTOb as shown in Fig. 2 [8, 12].

III. ANALYSIS OF THE REACTION TORQUE OBSERVER-BASED ROBUST FORCE CONTROLLER

This section briefly introduces the RTOb-based digital robust force controller and proposes a new stability analysis in the discrete-time domain.

*A. RTOb-based Robust Force Controller:*

The block diagram of the RTOb-based digital robust force controller is illustrated in Fig. 3. In this figure, $C_\tau$ represents the force control gain, $K_{env}$ and $D_{env}$ represent the environmental dynamics (i.e., the stiffness and damping of the environment, respectively), and $\tau_{ref}$ represents the force reference. The other parameters are same as defined earlier.

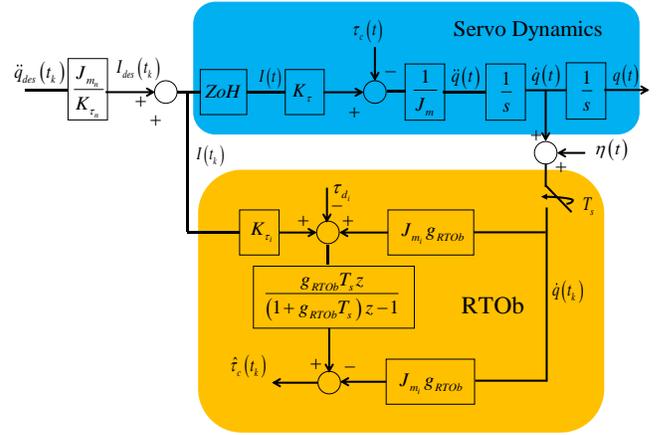

Fig. 2: Block diagram of the digital RTOb implemented by the Backward-Euler Integration method.

As shown Fig. 3, the robust force controller is synthesised using two observers in the inner- and outer- loop. In general, the robust force controller is synthesised by using $g_{DOb} = g_{RTOb}$ and assuming that $J_{m_n} = J_{m_i} = J_m$ and $K_{\tau_n} = K_{\tau_i} = K_\tau$. Let us show how this conventional design approach may lead to severe stability and performance problems in practice.

*B. Stability Analysis for the Robust Force Controller:*

The open-loop transfer function of the RTOb-based digital robust force controller can be derived from Fig. 3 as follows:

$$L_{RTOb}(z) = C_\tau J_{m_i} g_{RTOb} \beta \dfrac{T_s z}{z-1} \dfrac{(1+g_{DOb} T_s)z - 1}{(1+g_{RTOb} T_s)z - 1} \Phi(z) \quad (5)$$

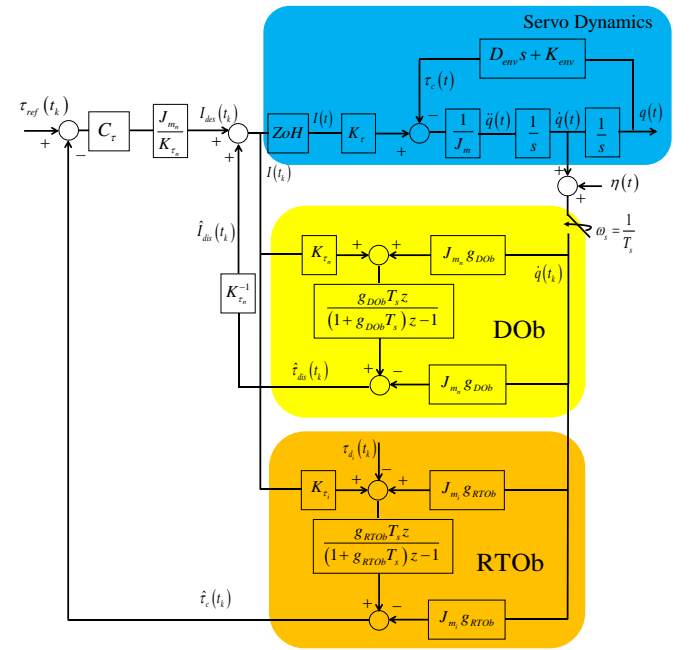

Fig. 3: Block diagram of the RTOb-based digital robust force controller implemented by the Backward-Euler Integration method.

where $\beta = (J_{m_n} K_{\tau_i})/(J_{m_i} K_{\tau_n})$, and $\Phi(z) = \Phi_n(z)/\Phi_d(z)$ in which

$$\Phi_n(z) = z^3 - \left(2e^{-\xi\omega_0 T_s}\cos(\omega_n T_s) + \delta e^{-\xi\omega_0 T_s}\text{sinc}(\omega_n T_s)\right)z^2 \quad (6)$$
$$+ \left(e^{-2\xi\omega_0 T_s} + 2\delta e^{-\xi\omega_0 T_s}\text{sinc}(\omega_n T_s)\right)z - \delta e^{-\xi\omega_0 T_s}\text{sinc}(\omega_n T_s)$$

$$\Phi_d(z) = z^3 - \left(2e^{-\xi\omega_0 T_s}\cos(\omega_n T_s) - \alpha g_{DOb}T_s e^{-\xi\omega_0 T_s}\text{sinc}(\omega_n T_s)\right)z^2 \quad (7)$$
$$+ \left(e^{-2\xi\omega_0 T_s} - \alpha g_{DOb}T_s e^{-\xi\omega_0 T_s}\text{sinc}(\omega_n T_s)\right)z$$

where $\delta = \alpha/\beta = (J_{m_i} K_\tau)/(J_m K_{\tau_i})$, $\omega_0 = \sqrt{K_{env}/m}$, $\xi = D_{env}/2\omega_0 m$, $\omega_n = \omega_0\sqrt{1-\xi^2}$ and $\text{sinc}(\omega_n T_s) = \sin(\omega_n T_s)/\omega_n T_s$.

Equation (5) shows that the open-loop transfer function of the digital robust force controller has an integrator which removes the steady-state error in force regulation. The open-loop transfer function has also a phase-lead/lag compensator which can be tuned by the bandwidths of the DOb and RTOb. A phase-lead compensator is synthesised using $g_{DOb} < g_{RTOb}$.

Equation (5) also shows that $L_{RTOb}(z)$ has the third order transfer function $\Phi(z)$ which significantly changes the stability and performance of the robust force controller when the control parameters, e.g., $\alpha$ and $\beta$, and environmental dynamics change. The poles of $\Phi(z)$ move towards the outside of the unit circle as $\alpha g_{DOb}T_s$ is increased. Since this may lead to a severe stability problem in practice, $\alpha$ and $g_{DOb}$ should have an upper bound. To relax the constraint on these design parameters, the sampling time of the robust force controller should be decreased. This is consistent with the robust stability and performance constraints of the inner-loop controller described in Section II. The zeros of the transfer function $\Phi(z)$ move towards the outside of the unit circle as $\delta e^{-\xi\omega_0 T_s}\text{sinc}(\omega_n T_s)$ increases, e.g., as the identified inertia (torque coefficient) is increased (decreased) in the design of the digital RTOb. In other words, the digital robust force controller has a non-minimum phase zero(s). The non-minimum phase zeros directly limit the bandwidth of the robust force controller, leading to poor performance. As the force control gain increases, the stability of the robust force controller deteriorates. To achieve good stability, smaller values of $\delta$ and $T_s$ should be used in the design of the RTOb-based digital robust force controller.

## IV. SIMULATIONS AND EXPERIMENTS

This section verifies the proposed stability analysis by simulations and experiments. Let us start with the design constraints in the inner-loop. Fig. 4 illustrates the frequency responses of the inner-loop's sensitivity and complementary sensitivity transfer functions. This figure shows that as $\alpha$ and $g_{DOb}$ are increased, the sensitivity function gets smaller at low-frequencies. In other words, the robustness against disturbances is improved. However, the waterbed effect is observed and the peaks of the sensitivity and complementary sensitivity functions increase to hold the constraint given in Eq. (2). Since the robust stability and performance deteriorate due to the waterbed effect, $\alpha$ and $g_{DOb}$ cannot be freely increased in the DOb synthesis.

Let us now consider the stability of the RTOb-based digital robust force controller. Figure 5 illustrates the root-loci of the robust force controller with respect to $C_\tau$ when different design parameters are employed in the DOb and RTOb synthesis. It is clear from this figure that the stability of the robust force controller deteriorates when i) DOb and RTOb are tuned using a higher value of $\delta$ (i.e., $\delta = 1$), and ii) the phase-lag controllers are employed by using $\alpha < 1$ and $g_{DOb} > g_{RFOb}$ in the inner- and outer- loop, respectively. Although the actual values of the inertia and torque coefficient are used in the RTOb synthesis, the digital robust force controller can suffer from non-minimum phase zeros. To improve the stability of the robust force controller, $\delta$ should be decreased and the digital robust force controller should be synthesised using the phase-lead controllers in the inner- and outer- loop as illustrated in Fig. 5.

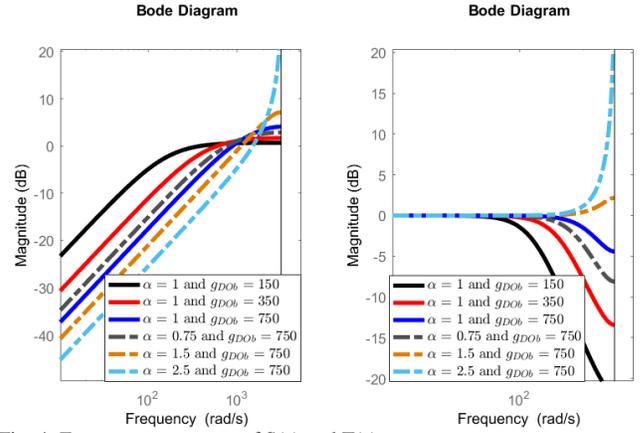

Fig. 4: Frequency responses of S(z) and T(z).

Last, let us experimentally verify the proposed stability analysis. The force control experiments were conducted for different values of the design parameter $\alpha$. As $\alpha$ is increased, the robust force controller cannot meet the design constraints proposed in the paper. Therefore, the stability of the digital robust force controller deteriorated as shown in Fig. 6.

## V. CONCLUSION

In this paper, a new stability analysis has been proposed for the RTOb-based digital robust force controllers in the discrete-time domain. The proposed stability analysis enables us to explain unexpected dynamic behaviours observed in practice, e.g., poor stability for higher values of $g_{DOb}$ and $J_{m_i}$. Moreover, it provides useful design tools to implement high-performance robust force controllers, e.g., i) the phase-lead/lag compensators in the inner- and outer- loop, and ii) minimum-phase robust force controller design by tuning $\delta$. Further studies should be conducted to obtain more practical design tools for the RTOb-based digital robust force controller.

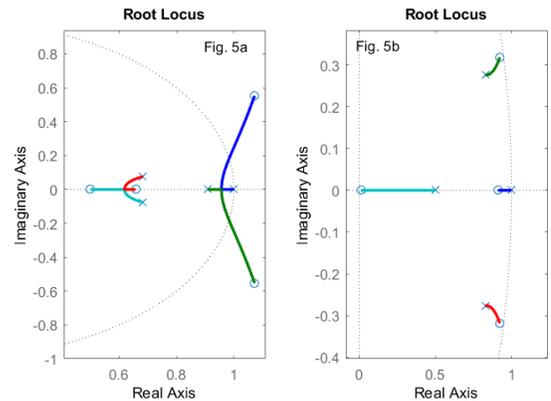

Fig. 5: Root loci of the digital robust force controller. a) $\alpha = 0.5$, $\delta = 1$, $g_{DOb} = 1000$ and $g_{RFOb} = 100$, b) $\alpha = 2$, $\delta = 0.1$, $g_{DOb} = 100$ and $g_{RFOb} = 1000$.

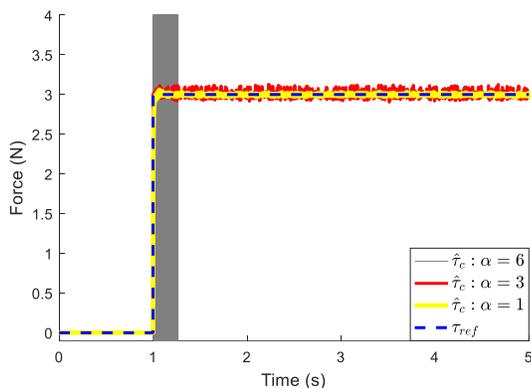

Fig. 6: Force control experiment when $g_{DOb} = 500$, $g_{RTOb} = 500$, $C_\tau = 0.6$, $T_s = 1 \times 10^{-3}$ and the different values of $\alpha$.